\documentstyle[prd,aps,preprint,epsfig]{revtex}
\tighten

\begin{document}
\draft
 
\pagestyle{empty}

\preprint{
\noindent
%\begin{minipage}[t]{3in}
%\begin{flushleft}
%\today \\
%\end{flushleft}
%\end{minipage}
\hfill
\begin{minipage}[t]{3in}
\begin{flushright}
LBNL--46203 \\
UCB--PTH--00/20 \\
hep-ph/0006296 \\
Sept. 2000 (revised)
\end{flushright}
\end{minipage}
}

\title{A possible hadronic excess in $\psi(2S)$ decay 
and the $\rho\pi$ puzzle}

\author{
Mahiko Suzuki
%\thanks{Work supported in part by the Director, Office of Energy
%Research, Office of High Energy and Nuclear Physics, Division of High
%Energy Physics of the U.S. Department of Energy under Contract
%DE--AC03--76SF00098 and in part by the National Science Foundation under
%grant PHY--95--14797.}
}
\address{
Department of Physics and Lawrence Berkeley National Laboratory\\
University of California, Berkeley, California 94720
}

%\thanks{Work supported by the Department of Energy under Contract
%DE--AC03--76SF00515.}

%\date{\today}
\maketitle

\begin{abstract}

We study the so-called $\rho\pi$ puzzle of the $\psi(2S)$ decay by 
incorporating two inputs; the relative phase between the 
one-photon and the gluonic decay amplitude, and a possible 
hadronic excess in the inclusive nonelectromagnetic decay rate of $\psi(2S)$. 
We look into the possibility that the hadronic excess in $\psi(2S)$
originates from a decay process of long-distance origin which is absent from 
the $J/\psi$ decay.  We propose that the amplitude of this additional process 
happens to nearly cancel the short-distance gluonic amplitude 
in the exclusive decay $\psi(2S)\rightarrow 1^-0^-$ and turn the sum 
dominantly real in contrast to the $J/\psi$ decay. We present general 
consequences of this mechanism and survey two models which 
might possibly explain the source of this additional amplitude.
 
\end{abstract}
%\pacs{}
%\newpage
\pagestyle{plain}
\narrowtext

\setcounter{footnote}{0}

\section{Introduction}
Absence of the $\rho\pi$ decay mode of $\psi(2S)$ has defied
a theoretical explanation for more than a decade\cite{Frankline}. 
The recent measurement by BES Collaboration\cite{Harris} has confirmed 
the absence of $\rho\pi$ with even a higher precision, setting its upper 
bound at a level of factor more than 60 below what one naively expects 
from the decay $J/\psi\rightarrow\rho\pi$. The measurement of other 
decay modes by BES\cite{Harris,BES} seems to rule out all possible 
resolutions for the $\rho\pi$ puzzle that have so far been proposed 
by theorists\cite{models}.
For instance, the large $\omega\pi$ branching contradicts the helicity 
suppression\cite{BLT0} with or without large intrinsic charm\cite{BLT}. 
A vector glueball near the $J/\psi$ mass, if it should exist, can enhance 
the $\rho\pi$ branching for $J/\psi$ relative to $\psi(2S)$\cite{Hou}. 
However, the magnitude of $B(J/\psi\rightarrow\rho\pi)$ 
is in line with expectation when we compare the 
$B(\rho\pi)/B(\omega\pi)$ with the inclusive ratio
$B(J/\psi\rightarrow ggg\rightarrow X)/B(J/\psi\rightarrow\gamma^*
\rightarrow X)$. What happens is not enhancement of $\rho\pi$ in 
$J/\psi$ but suppression of $\rho\pi$ in $\psi(2S)$.\footnote{
More comparison of experiment with models is found in Ref.\cite{Harris}.} 

Meanwhile the amplitude analysis of the $J/\psi$ decay revealed 
that the relative phase of the gluonic and the one-photon decay
amplitude is close to $90^{\circ}$ for all two-body decay channels so 
far studied; $1^-0^-$\cite{phase}, $0^-0^-$\cite{Kopke}, 
$1^-1^-$\cite{1-1-}, and $N\overline{N}$\cite{FENICE}. 
We show in this paper that the recent BES measurement in $J/\psi
\rightarrow 1^+0^-$ is also compatible with a large phase.

In contrast, the pattern of a large relative phase does not emerge 
for $\psi(2S)$.  Within experimental uncertainties, the relative 
phase is consistent with zero in the $1^-0^-$ 
decay\cite{BraatenTuan1,BraatenTuan2} and the $1^+0^-$ decay. 
This marked difference between $J/\psi$ and 
$\psi(2S)$ is another puzzle if the three-gluon decay is equally 
responsible for the strong decay of $J/\psi$ and $\psi(2S)$ 

There is one more experimental information relevant to the issue. That 
is the hadronic decay rate of $\psi(2S)$ which is normally attributed to
$\psi(2S)\rightarrow ggg$. When we compute with the current data the 
inclusive gluonic decay rate of $\psi(2S)$ by subtracting the cascade and 
the electromagnetic decay rate from the total rate, it is 60-70\%
larger, within experimental uncertainties, than what we expect from 
the short-distance gluonic decay alone. This excess hadronic branching 
in $\psi(2S)$ may suggest that something more occurs in the gluonic 
decay of $\psi(2S)$ than in the $J/\psi$ decay. 

In this paper we combine these informations together and search 
the origin of the marked difference between $J/\psi$ and $\psi(2S)$. 
While we should be apprehensive about experimental errors at present, they 
might give us a clue to a solution of the $\rho\pi$ puzzle. In Section II,
prompted by the experimental observation in the $0^-0^-$, $1^-0^-$, $1^-1^-$,
and $N\overline{N}$ channels, we postulate universality of 
the large relative phase between the gluon and the photon decay 
amplitudes. Specifically, the gluonic decay amplitude acquires 
a large phase while the photon amplitude is real.  We point out that
a large phase is consistent with new BES data in $J/\psi\rightarrow 1^+0^-$.
Further progress in the BES analysis in this channel will shed more light.
We turn to $\psi(2S)$ in Section III. The decay branching fractions
of $\psi(2S)\rightarrow 1^-0^-$ clearly show suppression of the gluon 
amplitude and favor a small relative phase between the gluon and the photon 
amplitude. We point out that a small phase is more likely in 
$\psi(2S)\rightarrow 1^+0^-$ too. Taking the possible excess in the 
inclusive hadronic decay rate of $\psi(2S)$ seriously, we propose 
that this excess is related to both the suppression 
and the small relative phase of the $1^-0^-$ amplitude. 
Our proposition is that an additional decay process
generating the excess should largely cancel the short-distance
gluon amplitude in the exclusive decay into $1^-0^-$ and that the resulting 
small residual amplitude is not only real but also destructively 
interferes with the photon amplitude. In Section IV, we first 
present general consequences of the destructive interference.  
We then examine two scenarios which may possibly generate 
the excess inclusive gluonic decay. One is the contribution of the virtual 
$D\overline{D}$ intermediate state. The other is a resonance, 
a glueball or a four-quark resonance, near the $\psi(2S)$ mass. 
Though neither idea is novel nor highly appealing, they seem to be 
among a very few possibilities that have not yet been ruled out 
by experiment.

\section{Phases of $J/\psi$ decay amplitudes}

The relative phase between the gluon and the photon amplitude in the
decay $J/\psi\rightarrow 1^-0^-$ has been analyzed with broken 
flavor SU(3) symmetry\cite{phase} including the $\rho$-$\omega$ mixing.  
All analyses clearly show that the relative phase should be very large 
and not far from $90^{\circ}$ with experimetnal uncertainties. The SU(3) 
analysis was made also for the $0^-0^-$ modes\cite{Kopke} and the
$1^-1^-$ modes\cite{1-1-} for which leading gluon amplitude is SU(3) 
violating. The relative phase were found to be equally large for these 
modes. Furthermore comparison of the electromagnetic form factors 
in the timelike region with the $J/\psi$ decay branching fractions revealed 
that the relative phase is very close to $90^{\circ}$ in the $N\overline{N}$
decay channels too\cite{FENICE}.  A question arises as to whether 
this large relative phase is universal to all decays of 
$J/\psi$ or not. There is no persuasive theoretical answer to it at 
present.\footnote{Some attempt was recently made to argue in favor
of universal large phases\cite{Weyers}.}

In addition to those two-body channels already analyzed, the recent BES
measurement\cite{BES} on $J/\psi\rightarrow 1^+0^-$,
\begin{eqnarray}
      B(J/\psi\rightarrow K_1^{\pm}(1400)K^{\mp}) &=&
                         (3.8\pm 0.8 \pm 1.2)\times 10^{-3}, \\ \nonumber
      B(J/\psi\rightarrow K_1^{\pm}(1270)K^{\mp}) &<&
                          3.0\times 10^{-3},\;\;\; 90\%\;C.L.
\end{eqnarray}
is relevant to this issue. We examine here these branching fractions
together with the $b^{\pm}\pi^{\mp}$ branching fraction,
$B(J/\psi\rightarrow b^{\pm}\pi^{\mp})=(3.0\pm 0.5)
\times 10^{-3}$\cite{PDG}.\footnote{A
previous analysis\cite{axial} of these $J/\psi$ decay modes
assumed a zero relative phase and used on a preliminary value of the
upper bound on $B(J/\psi\rightarrow K_1^{\pm}(1270)K^{\mp})$.
Therefore the $J/\psi$ analysis of Ref.\cite{axial} should be
disregarded. However, the analysis of the $\psi(2S)\rightarrow 1^+0^-$
of Ref.\cite{axial} remains valid.}

   Since $K_1(1270)$ and $K_1(1400)$ are superpositions of $K_{A}$ and
$K_{B}$ of the $1_A^{++}$ and the $1_B^{+-}$ octet, respectively,
\begin{eqnarray}
      K_1^+(1400) &=& K_A^+\cos\theta + K_B^+\sin\theta, \\ \nonumber
      K_1^+(1270) &=& - K_A^+\sin\theta + K_B^+\cos\theta,
\end{eqnarray}
with $\theta\approx 45^{\circ}$\cite{theta}, we can parametrize the
three decay amplitudes in terms of the gluon amplitude $a_1$ of the
$1_B^{+-}$ octet and the photon amplitudes $a_{\gamma A/B}$
of $1_A^{++}$ and $1_B^{+-}$;
\begin{eqnarray}
      A(b_1^+\pi^-) &=& a_1 +\sqrt{1/5}a_{\gamma B}, \nonumber \\
  A(K_1^+(1270)K^-) &=& (a_1 +\sqrt{1/5}a_{\gamma B})\cos\theta
                    - a_{\gamma A}\sin\theta,         \nonumber \\
  A(K_1^+(1400)K^-) &=& (a_1 + \sqrt{1/5}a_{\gamma B})\sin\theta
                    + a_{\gamma A}\cos\theta.  \label{1+}
\end{eqnarray}
Since there are two independent helicity amplitudes (or $s$- and
$d$-waves) for $1^+0^-$, we should use this parametrization
separately for the $s$-wave and the $d$-wave amplitudes.
The three branching fractions can be fitted with
\begin{eqnarray}
      |a_1| &>& |a_{\gamma A}|\approx |a_{\gamma B}|, \nonumber \\
 \arg(a_1^*a_{\gamma A})&\approx &\arg(a_1^*a_{\gamma B})\approx 90^{\circ}.
\end{eqnarray}
If the ratio $\Gamma(J/\psi\rightarrow\gamma\rightarrow 1^+0^-)/\Gamma(J/\psi
\rightarrow ggg\rightarrow 1^+0^-)$ is comparable in magnitude to
$\Gamma(J/\psi\rightarrow\gamma\rightarrow X)/\Gamma(J/\psi
\rightarrow ggg\rightarrow X)\simeq 1/5$, we should expect
that $|a_{\gamma A/B}|\approx 0.7|a_1|$\cite{axial}. For $a_{\gamma A}
= a_{\gamma B}= \pm 0.7i a_1$ and $\theta\approx 45^{\circ}$, the
ratios of the branching fractions prior to the phase space corrections,
denoted by $B_0$, take values as
\begin{equation}
B_0(b^{\pm}\pi^{\mp}):B_0(K_1^{\pm}(1270)K^{\mp}):B_0(K_1^{\pm}(1400)K^{\mp})
                   \simeq 1 : 0.5 : 0.9.
\end{equation}
While the inequality $B(K_1^{\pm}(1270)K^{\mp}) <
B(K_1^{\pm}(1400)K^{\mp})$ can easily be realized by a wide
range of parameter values, the other inequality $B(b^{\pm}\pi^{\mp})
< B(K_1^{\pm}(1400)K^{\mp})$ is a little tight.
If we allow $|a_{\gamma A/B}|$ larger than $0.7|a_1|$ and/or
increase the value of $\theta$, however, the current central values
of the branching fractions are consistent with the large phase 
hypothesis. We should also point out that if the SU(3) breaking 
correction is made by the meson wavefunctions $(f_Pf_A)^2$, 
it is likely to enhance $B(K_1^{\pm}(1400)K^{\mp})$ over
$B(b^{\pm}\pi^{\mp})$ to the direction in favor of the large phase fit.

If we leave $a_{\gamma A/B}$ unrestricted in magnitude and phase,
a triangular relation holds for the amplitudes as 
\begin{equation}
    A(K_1^{\pm}(1270)K^{\mp})\cos\theta 
  + A(K_1^{\pm}(1400)K^{\mp})\sin\theta
    =A(b^{\pm}\pi^{\mp}). \label{sumrule}
\end{equation}
Determination of the $s$-to-$d$ wave ratio of the amplitudes and
further study of $B(J/\psi\rightarrow K_1^{\pm}(1400)K^{\mp})$
will eventually resolve the composition of amplitudes and test
the large phase hypothesis in $1^+0^-$.
As it was pointed out previously\cite{axial}, it is also
important to resolve the discrepancy between $B(b^{\pm}\pi^{\mp})$
and $2B(b^0\pi^0)$\cite{PDG}, which theory predicts to be equal.

To summarize the experimental situation of the two-body $J/\psi$ decay
amplitudes, the existing data strongly favor large relative phases 
close to $90^{\circ}$ between the gluon and the photon decay 
amplitudes for $1^-0^-$, $0^-0^-$, $1^-1^-$, and $N\overline{N}$, 
and are consistent with a large phase for $1^+0^-$. 

What does theory say about these relative phases ?
In the perturbative picture, the gluonic decay of $J/\psi$ proceeds 
as depicted in Fig.1a. The inclusive decay rate is computed with the 
gluons placed on mass shell. In contrast, the photon being far off shell,
no corresponding on-shell intermediate state appears in the perturbative 
diagrams of the photon amplitude (Fig.1b). 
Although perturbative QCD (pQCD) is a good description of inclusive
charmonium decays, it is questionable whether it works for two-body
decay channels of charmonia. To be specific, the pQCD prediction 
of the asymptotic pion form factor\cite{BL},
\begin{equation}
       F_{\pi}(q^2) \simeq 16\pi\alpha_s(q^2)f_{\pi}^2/q^2,
\end{equation}
has not been reached at the $J/\psi$ mass\cite{Nussinov}. 
Furthermore, the helicity suppression argument of 
pQCD fails for the $\omega\pi$ decay channel. 
Though it is tempting, therefore, we cannot argue for the large 
relative phases on the basis of perturbative diagrams. Whether or not
these large relative phases are universal to all two-body decay modes 
of $J/\psi$ must be determined by experiment. Despite lack of a good
theoretical argument at present, we suspect nonetheless that the universal 
large phases so far found are not an accident.

If the relative phases are close to $90^{\circ}$, it is more likely that
the photon decay amplitudes are real and consequently the gluon decay 
amplitudes are imaginary. The reason is as follows: In order for the
photon decay amplitude to have a substantial phase, the final $q\overline{q}$ 
created by the virtual $\gamma$ should have a large absorptive part of 
a long-distance origin. This can happen if there should be a relatively sharp
$q\overline{q}$ resonance just around $J/\psi$. More likely is that
many resonances exist below $J/\psi$ as in the
vector-meson-dominance scenario, or that as in the dual resonance 
model, many increasingly broader resonances appear all the way to high 
energies\cite{KK}. In the former case, only the tails of low-lying 
resonances contribute to the real part. In the latter case, a few nearby
broad resonances can contribute to the imaginary part, but they are 
outnumbered by many more resonances below and above $J\psi$ 
that contribute to the real part. In comparison, we have less insight in 
hadronization dynamics of the gluon decay.  

Motivated by the results in the amplitude analyses of the two-body of 
$J/\psi$ decays, we make two postulates:\\
\indent (1) The relative phases between the gluon and the photon decay 
amplitudes are universally large for all two-body decays of $J/\psi$.
The photon decay amplitudes are predominantly real and consequently the
gluon decay amplitudes are imaginary. \\ 
\indent (2) The same pattern holds for $\psi(2S)$ decay as well.\\
These are the starting assumptions of our analysis that follows.
  
\section{$\psi(2S)$ decay}

\subsection{Relative phase from experiment}

The only large energy scale involved in the three-gluon decay of charmonia is
the charm quark mass $m_c$. Whether one accepts the argument of the universal 
large phase in exclusive channels or not, therefore, one would naively 
expect that the corresponding phases should not be much different between 
the $J/\psi$ decay and the $\psi(2S)$ decay.  However, experimental data
so far available show that the phases are small at least in some two-body
decay modes of $\psi(2S)$.  The strongest evidence is 
in the decay $\psi(2S)\rightarrow 1^-0^-$, which includes the puzzling 
$\psi(2S)\rightarrow \rho\pi$. In the case that the final $0^-$ meson
is an octet, we can parametrize the $\psi(2S)\rightarrow 1^-0^-$ decay 
amplitudes with the SU(3) singlet amplitude $a_1$, the SU(3) breaking 
correction $\epsilon$ due to $m_s-m_{ud}$, and the photon 
amplitude $a_{\gamma}$. The corresponding amplitudes, 
$b_1$ and $b_{\gamma}$, are introduced for the $0^-$ singlet. 
For the $\phi$-$\omega$ mixing, we assume the nonet scheme. The 
parametrization of the amplitudes\cite{Seiden} 
is listed in Table I for the decay 
modes so far studied in experiment\cite{Harris}. Since the analyzed 
channels are limited and the uncertainties in their branching 
fractions are still large, we are unable to perform a meaningful 
$\chi^2$ fit at present. Therefore, we present only fits to 
the central values by referring to Table I. 

First of all, if we ignored the photon amplitude $a_\gamma$, 
we would obtain $B(K^{*0}\overline{K}^0+ c.c.) =
B(K^{*\pm}K^\mp)$, which contradicts with experiment, $(0.81\pm 0.24\pm 
0.16)\times 10^{-4}$ {\it vs} $< 0.30\times 10^{-4}$. The large splitting 
between these branching fractions requires that $a_{\gamma}$ be comparable 
to $a_1$. If we set $a_1$ and $\epsilon$ to zero,  
we would obtain up to phase space corrections
\begin{equation}
            B_0(\omega\pi)/B_0(K^{*0}\overline{K}^0 + c.c.)
             = 9/8  \label{8/9}
\end{equation}
in contradiction with the measurement, $(0.38\pm 0.17 \pm 0.11)/(
0.81\pm 0.24\pm 0.16)$.\footnote{
It is possible that the SU(3) breaking in the strange and nonstrange 
meson wavefunctions $(f_{\pi}f_{\rho}/f_Kf_{K^*})^2$ may be responsible for
part of the discrepancy.} In order to come closer to this ratio of 
the measured values, a large constructive interference should 
occur in $K^{*0}\overline{K}^0$, that is, the relative phase 
must be small between $a_1+\epsilon$ and $-2a_{\gamma}$. 
Then, assuming that $a_1$ and $\epsilon$ have a common phase, we have
a large destructive interference between $a_1$ and $a_{\gamma}$ for both 
$\rho\pi$ and $K^{*\pm}K^{\mp}$ in agreement with experiment.\footnote{
This main feature of the fit to the $\psi(2S)\rightarrow 1^-0^-$ amplitudes 
is found in the earlier paper by Chen and Braaten\cite{BraatenTuan1} 
and, in particular, in the paper by Tuan\cite{BraatenTuan2}.} This solves 
the $\rho\pi$ puzzle and explains also the missing of the 
$K^{*\pm}K^{\mp}$ mode in experiment.
 
With these qualitative observations in mind, we have fitted 
to the central values of the observed branching fractions and 
then have computed with those parameter values the branching fractions 
of the modes for which only the upper bounds have been determined. 
In Table I we have listed the fit with $\delta \equiv -
\arg(a_1^*a_{\gamma}) = 0^{\circ}$ and the large phase fit with $\delta 
= \pm 90^{\circ}$ for comparison. When the central values of
$B(\omega\pi)$ and $B(K^{*0}\overline{K}^{0}+ c.c.)$ are fitted 
with $\delta = 0^{\circ}$, the ratio of the photon and the gluon
amplitude turns out to be
\begin{equation}
     a_{\gamma}/(a_1+\epsilon) \simeq -0.76. \label{suppression}
\end{equation}
For comparison, $a_{\gamma}/(a_1+\epsilon)\simeq 0.14$ in the case 
of $J/\psi$. We would expect $|a_{\gamma}/a_1|\approx 0.22$ if the 
ratio of $\Gamma(\psi(2S)\rightarrow\gamma^*\rightarrow 1^-0^-)$ to 
$\Gamma(\psi(2S)\rightarrow ggg\rightarrow 1^-0^-)$ is roughly 
equal to $\Gamma(\psi(2S)\rightarrow\gamma^*\rightarrow X)
/\Gamma(\psi(2S)\rightarrow ggg\rightarrow X)$. Since experiment
shows that $\Gamma(\omega\pi)/\Gamma(l^+l^-) (\propto |a_{\gamma}|^2)$ 
is about the same for $J/\psi$ and $\psi(2S)$, the large number 
in Eq.(\ref{suppression}) results from strong suppression of 
the total gluonic amplitude $a_1+\epsilon$ in $\psi(2S)$. 
As the value of $\epsilon$ is varied in the range of $|\epsilon/a_1|
< 1/3$, the value of $B(\rho^{\pm}\pi^{\mp})$ varies between 0 and 
0.04$\times 10^{-4}$. The values for $B(\rho^{\pm}\pi^{\mp})$ and 
$B(K^{*\pm}K^{\mp})$ can be increased if we stretch 
within the experimental uncertainties of $B(\omega\pi)$ and 
$B(K^{*0}\overline{K}^0 + c.c.)$. In contrast, 
the fit with $\delta =\pm 90^{\circ}$ overshoots the upper 
bound on $B(K^{*\pm}K^{\mp})$ and, if $|\epsilon| < \frac{1}{3}|a_1|$, 
the upper bound on $B(\rho^{\pm}\pi^{\mp})$ very badly.  A fit  
with $\delta=\pm 90^{\circ}$ is virtually impossible even with 
experimental uncertainties unless $|\epsilon|\gg|a_1|$.  
We thus conclude that 
the relative phase between $a_1$ and $-a_{\gamma}$ should be small  
in $\psi(2S)\rightarrow 1^-0^-$ contrary to the $J/\psi$ decay.

Though it is less conclusive, a small phase seems to be favored in 
the $1^+0^-$ decay of $\psi(2S)$ too. It is conspicuous in 
experiment\cite{BES} that the $K_1^{\pm}(1400)K^{\mp}$ 
mode is strongly suppressed relative to the $K_1^{\pm}(1270)K^{\mp}$:
\begin{eqnarray}
  B(\psi(2S)\rightarrow K_1^{\pm}(1270)K^{\mp})
      & =&(10.0\pm 1.8\pm 2.1)\times 10^{-4}, \\ \nonumber
  B(\psi(2S)\rightarrow K_1^{\pm}(1400)K^{\mp}) &<& 3.1 \times 10^{-4}.
                         \label{1+0-}
\end{eqnarray}
$B(\psi(2S)\rightarrow b^{\pm}\pi^{\mp})$ is half way between them\cite{BES}:
\begin{equation}
B(\psi(2S)\rightarrow b^{\pm}\pi^{\mp}) = 
          (5.2 \pm 0.8 \pm 1.0)\times 10^{-4}.
\end{equation}
We can use Eq. (4) as the parametrization of 
$\psi(2S)\rightarrow 1^+0^-$. First of all, if $a_1$ dominated
over $a_{\gamma A/B}$, we would have $B(b^{\pm}\pi^{\mp}) \simeq
2B(K_1^{\pm}(1270)K^{\mp})\simeq 2B(K_1^{\pm}(1400)K^{\mp})$ for 
$\theta \simeq 45^{\circ}$ in disagreement with experiment. Just as 
in $\psi(2S)\rightarrow 1^-0^-$, $|a_1|$ is comparable to  
$|a_{\gamma A/B}|$. Next, the strong suppression of $K_1^{\pm}(1400)
K^{\mp}$ relative to $K_1^{\pm}(1270)K^{\mp}$ can be realized only when 
$a_1+\sqrt{1/5}a_{\gamma B}$ interferes destructively with $a_{\gamma A}$.
Therefore, the relative phase between $a_1$ and $a_{\gamma A/B}$
must be small modulo $\pi$. To obtain $B(K_1^{\pm}(1270)K^{\mp})\approx
2B(b^{\pm}\pi^{\mp})$, we need $a_1+\sqrt{1/5}a_{\gamma B}
\approx -a_{\gamma A}$. The allowed range of the amplitude ratios was 
plotted in Ref.\cite{axial} by choosing all amplitudes as relatively real 
and assuming tentatively the s-wave decay for phase-space corrections. 
Though it is not impossible to fit the three branching fractions with 
$\theta\approx 90^{\circ}$, we must have $a_1\simeq 0$ and 
$\sqrt{1/5}a_{\gamma B} \simeq -a_{\gamma A}$ in that case. 

To summarize for $\psi(2S)$, the data on $1^-0^-$ virtually excludes the
possibility of a small 
phase between  $a_1$ and $a_{\gamma}$. A fit to $1^+0^-$ has more room 
when the relative phase is small. There is no evidence for that 
the relative phase must be large in $\psi(2S)$. For some reason, 
a large relative phase does not seem to occur in the $\psi(2S)$ decay. 
We ask what causes this marked difference between $J/\psi$ and $\psi(2S)$ 
when we postulate the universal large phase for $\psi(2S)$ as well as
for $J/\psi$.

\subsection{Excess hadronic rate in inclusive hadronic decay}
   It has been noticed that when one computes the inclusive hadronic 
decay rate of $\psi(2S)$ through $ggg$ by subtracting the rates of the  
cascade and electromagnetic decays from the total decay rate, 
it is substantially larger than what we expect from an extrapolation 
of $J/\psi$. The number with a conservative error estimate based on the 
listings of {\it Reviews of Particle Physics}\cite{PDG} is 
\begin{equation}
\frac{B(\psi(2S)\rightarrow ggg+gg\gamma)}{B(J/\psi\rightarrow ggg+gg\gamma)}
    = 0.23 \pm 0.07, \label{R1}
\end{equation}
which should be compared with 
\begin{equation}
 \biggl(\frac{\alpha_s(\psi(2S))}{\alpha_s(J/\psi)}\biggr)^3
    \frac{B(\psi(2S)\rightarrow\l^+\l^-)}{B(J/\psi\rightarrow\l^+\l^-)}
    = 0.134\pm 0.034. \label{R2}
\end{equation}
Smaller errors ($0.226\pm 0.052$ {\it vs} $0.141\pm 0.012$) have been 
attached in a recent literature\cite{GuLi} with a different error estimate. 
We would expect that the two numbers should be equal to each other since
the wavefunctions at origin appear in common in Eqs.(\ref{R1}) 
and (\ref{R2}). The discrepancy of 60-70\% between them
alarms us particularly because all numbers involved have been repeatedly 
measured over many years.\footnote{The author learned that BES 
collaboration is considering a different determination of the 
cascade decay branchings\cite{Harrisp}.}

In comparison we find no similar excess in $\Upsilon(2S)$ though 
experimetnal undertaintues are large. In terms of the ratio of 
branching ratios, $\overline{B}(ggg + 
gg\gamma)\equiv B(\Upsilon\rightarrow ggg+gg\gamma)/\alpha_s(\Upsilon)^3
B(\Upsilon\rightarrow \mu^+\mu^-)$, three $\Upsilon$'s are more in line: 
\begin{equation}     
     \overline{B}(ggg+gg\gamma)= \left\{ \begin{array}{cc}
                          (4.5 \pm 0.2)\times 10^3, &  \Upsilon(1S)\\
                          (4.9 \pm 0.9)\times 10^3, &  \Upsilon(2S)\\
                          (4.0 \pm 0.4)\times 10^3, &  \Upsilon(3S).
                \end{array} \right.,
\end{equation}
where the total leptonic branching for $\Upsilon(3S)$\footnote{
The branching to $\Upsilon(2S)\gamma\gamma$ quoted in \cite{PDG} is sum of
the cascade decay branchings $\chi_{bJ}(2P)\gamma\rightarrow\Upsilon(2S)
\gamma$ in view of the $\gamma\gamma$ invariant mass spectrum\cite{Butler} 
and also of its magnitude ($>B(\Upsilon(2S)\pi^0\pi^0)$). 
It is counted in the radiative 
decay branchings separately listed in \cite{PDG}.} has been substituted
with three times $B(\mu^+\mu^-)$, the only quoted leptonic branching.
It appears that the excess in $\overline{B}(ggg+gg\gamma)$ is unique 
to $\psi(2S)$. However, this excess in the inclusive rate has not shown up
in the rates of the exclusive channels so far measured. In fact, the ratio
$B(\psi(2S)\rightarrow h)/B(J/\psi\rightarrow h)$ scatter around the 
expected value ($\approx$ 13-14\% of Eq.(\ref{R2})), which was often
called the {\it 14\% rule}. Some remarks should be in order on it.

First of all, the 14\% rule is largely violated in many of two-body and 
quasi-two-body channels, as we recently learned in the BES data\cite{Harris}.
The $\rho\pi$ channel is an extreme case. For multihadron channels, 
there are actually not so many modes that are available for testing 
the 14\% rule.\footnote{Gu and Li\cite{GuLi} lumped multihadron modes 
together and compared between $J/\psi$ and $\psi(2S)$. Then the number is 
dominated by three modes, $\pi^+\pi^-\pi^0$, $2(\pi^+\pi^-)\pi^0$ and 
$3(\pi^+\pi^-)\pi^0$, which happen to be below 14\%. The mode 
$\pi^+\pi^-\pi^0$ is actually $\rho\pi$ and its nonresonant content is 
consistent with zero\cite{PDG}. 
More revealing are the ratios of individual modes.}
In Table II, we have tabulated the ratios for the modes not listed in 
\cite{Harris} but available for comparison. We see that the ratios scatter 
rather widely above and below 14\% with some tendency of being smaller than 
14\%, but with fairly large experimental uncertainties. It is important 
to notice that the branching fractions of all modes in Table II 
add up to no more than 15\% of the total gluonic decay branching of $J/\psi$.
Indeed, only one charge state has been available for comparison from
each of $5\pi$, $7\pi$, $2\pi K\overline{K}$, and $N\overline{N}n\pi$.   
We have not yet seen comparison of the rest. The so-called 14\% rule 
is based on very limited number of decay modes. It is 
premature to preclude the hadronic excess with the data of
multihadron exclusive channels.

If future experiment shows that the hadronic excess in $\psi(2S)$ 
is real, it may have something to do 
with the $\rho\pi$ puzzle\cite{GuLi} and with the abrupt change of the 
relative phase of amplitudes from $J/\psi$ to $\psi(2S)$. 
A process responsible for the excess inclusive hadron rate can interfere 
with the short-distance gluon decay in exclusive modes. If its amplitude 
makes a large destructive interference with the three-gluon amplitude 
in $\psi(2S)\rightarrow 1^-0^-$ and if the sum is nearly real and comparable
to the photon amplitude in magnitude, our puzzle can be solved. 
We shall look into possible sources of this rate excess in the following.

\section{Additional hadronic amplitude in $\psi(2S)$}

\subsection{General consequences}
If the origin of the problem is in the interference of an unknown
additional process with the short-distance gluon decay of $\psi(2S)$, 
we expect a general pattern of correlation between the decay 
angular distribution and suppression or enhancement.

The decay angular distribution for two final hadrons is generally of the form,
\begin{equation}
    d\Gamma/d\Omega \propto 1 + a\cos^2\theta,\;\;\;(|a|\leq 1)
\end{equation}
where $\theta$ is the polar angle measured from the $e^+e^-$ 
beam direction. For $1^-0^-$ and $0^-0^-$ decays, the value of 
$a$ is constrained kinematically to +1 and $-1$, respectively, 
while it is determined dynamically by the helicity content, 
$\pm 1$ or 0, of the final state in other decays.
In $1^-0^-$ and $0^-0^-$, therefore, any additional amplitude has 
the same angular dependence as the three-gluon and the photon 
amplitude irrespective of its origin. Consequently a high degree 
of destructive or constructive interference with an additional 
amplitude is possible in these decays. 
Observation of the strongest suppression in the $1^-0^-$ mode 
is consistent with this pattern\cite{Harris}. We expect that the decay 
rates of $\psi(2S)\rightarrow 0^-0^-$ may also be quite different from those
of $J/\psi\rightarrow 0^-0^-$. In terms of
$\overline{B}(0^-0^-)\equiv B(0^-0^-)/\alpha_s^3B(\mu^+\mu^-)$,
the current data\cite{PDG} give
\begin{eqnarray}
     \overline{B}(\pi^+\pi^-)& =&\left\{ \begin{array}{ll}
                                0.15\pm 0.02 & \mbox{for $J/\psi$} \\
                                0.8 \pm 0.5 &  \mbox{for $\psi(2S)$}
                                \end{array} \right.           \\ \nonumber
     \overline{B}(K^+K^-) &=& \left\{ \begin{array}{cc}
                                0.24\pm 0.03 & \mbox{for $J/\psi$}  \\
                                0.94 \pm 0.66 &\mbox{for $\psi(2S)$}
                                \end{array} \right.
\end{eqnarray}
Within the large experimental uncertainties we see a hint of
large constructive interference in $\psi(2S)\rightarrow
0^-0^-$.  In contrast, in other processes  an additional 
amplitude and the three-gluon amplitude have different angular 
distributions in general. A large interference can occur only when
dynamical mechanisms of two processes are similar. Otherwise it 
should be a result of a high degree of accident. When a large disparity 
is observed between corresponding two-meson decay rates of 
$J/\psi$ and $\psi(2S)$, therefore, the decay angular distribution of 
this channel will also be very different between $J/\psi$ 
and $\psi(2S)$. This will give a good test of the idea of 
interference with an additional amplitude.

The other consequence is in multibody final states. Since the 
additional process enhances the inclusive rate, a large number of 
exclusive decay channels should receive enhancement rather
than suppression. When there are many hadrons in the final state, 
chance of interference between amplitudes of different decay 
mechanism is much smaller because of difference in subenergy 
dependence and event topology. Therefore enhancement will not be dramatic.
While we have not yet seen such enhancemnet in Table II, we expect 
that the branching fraction tend to be enhanced in  
many nonresonant multibody channels of $\psi(2S)$ 
relative to $J/\psi$.
 
Where does the additional amplitude possibly come from ?
There are few options left in modifying charmonium physics
radically. Since aspects of perturbative QCD have been well understood,
we are bound to look for the origin of the problem in long-distance
physics of one kind or another.

\subsection{$\psi(2S)\rightarrow D\overline{D}\rightarrow$ hadrons}

One unique feature of $\psi(2S)$ is a close proximity of its mass to the
$D\overline{D}$ threshold. The $\psi(2S)$ mass is only
43 MeV (53 MeV) below $D^0\overline{D}^0$ ($D^+D^-$), while
$\Upsilon(3S)$ is 200 MeV away from the $B\overline{B}$ threshold.
Can the small energy difference\footnote{This was brought to the 
author's attention by J.L. Rosner\cite{JLR}.} between $\psi(2S)$ and
$D\overline{D}$ have anything to do with the excess ?
It may happen that $\psi(2S)$ picks up a light quark pair
through soft gluons and dissociates virtually into $D\overline{D}$,
which in turn annihilate into light hadrons. (See Fig.2.)
The dominant process of the $D\overline{D}$ annihilation is
through $c\overline{c}$ annihilation through a single hard gluon.
The small energy denominator enhances creation of virtual
$D\overline{D}$ while $p$-wave creation compensates the enhancement.
It is difficult, actually nearly impossible, to give a reliable computation
of this sequence. Deferring estimate of the rate to future,
we shall comment here only on whether the $D\overline{D}$ contribution 
can have a final-state interaction phase large enough to cancel 
the perturbative gluon amplitude or not.

Since the $D\overline{D}$ intermediate state is above the $\psi(2S)$
mass, a phase of amplitude must come from the subsequent annihilation
of $c\overline{c}$ and thereafter. After an energetic light quark pair 
$q\overline{q}$ is created from $c\overline{c}$, each of $q\overline{q}$ 
picks up a soft light quark from the light quark cloud of $D\overline{D}$ 
to form mesons. (See Fig.3a.) Kinematically, this step of the hadron 
formation process is quite different from that of the timelike 
electromagnetic form factor of a meson in which energetic light quarks 
pick up collinear quarks created by a hard gluon. (See Fig.3b.) In our 
case color-dipole moment is large for all pairs of quarks\cite{BJ}.
Furthermore, the {\it c.m.} energy of a hard quark in one meson and 
a soft quark in the other meson is in the low energy resonance region, 
\begin{equation}
            \sqrt{s} = O(\sqrt{2\Lambda_{QCD}m_c}) < 1{\rm GeV}. 
\end{equation}
Therefore, one cannot argue that final-state interactions should be small 
between final mesons. We expect that there is a good chance for the amplitude 
of $\psi(2S)\rightarrow D\overline{D}\rightarrow mesons$ to acquire
a substantial final-state interaction phase. Weakness in this argument 
is that the phase can be large but need not be large.

The idea of the virtual $D\overline{D}$ dissociation actually has
some common feature with that of the ``higher Fock component'' of
charmonia\cite{Kroll,Braaten}. The $D\overline{D}$ state can be
viewed as part of the four-quark Fock space of $\psi(2S)$.
The higher Fock component was proposed as an additional contribution
to $J/\psi\rightarrow 1^-0^-$ to solve the $\rho\pi$ puzzle\cite{Braaten}.
It was argued that it is more significant in $J/\psi$ than in $\psi(2S)$. 
As we have emphasized, however, there is nothing anomalous 
about $J/\psi\rightarrow 1^-0^-$.

\subsection{$\psi(2S)\rightarrow$resonance$\rightarrow$ hadrons}
The second idea is a twist of an old one: A noncharm resonance may exist near
the $\psi(2S)$ mass and give an extra contribution to the hadronic decay rate.
A glueball was proposed earlier at the $J/\psi$ mass to boost the
$\rho\pi$ decay rate of $J/\psi$\cite{Hou}.  However, we now want it near
$\psi(2S)$ not near $J/\psi$.  We look into the possibility that some 
resonance around the $\psi(2S)$ mass, a glueball or four-quark,
destructively interferes with the perturbative $\psi(2S)\rightarrow
ggg\rightarrow 1^-0^-$ decay. Admittedly, the idea is {\it ad hoc} 
and there is some difficulty aside from unnaturalness.
  
  Light-quark resonances of high mass ($\sim 3.7$ GeV) and low
spin are normally too broad to be even recognized as resonances. 
Four-quark resonances may be an alternative if they exist at all. The mass 
of $3.7$ GeV is normally considered as too high for the lowest vector 
glueball. An excited glueball state of $J^PC=1^{--}$ serves our purpose.
Whatever its origin is, let us introduce here such a resonance,
call it $R$, and see its consequences.

  In order for $\psi(2S)$ to decay through $R$ as strongly as through 
three gluons, the coupling $f$ of $R$ to $\psi(2S)$ defined by 
$-fm_R\psi_{\mu}R^{\mu}$ must be large enough. The $\psi(2S)$-$R$ mixing
at the $\psi(2S)$ mass is given by
\begin{equation}
            \varepsilon \simeq\frac{f}{\Delta m-i\Gamma_R},
\end{equation}
where $\Delta m = m_R-m(\psi(2S))$ and $\Gamma_R$ is the total width 
of $R$. It leads to $\Gamma(\psi(2S)\rightarrow R\rightarrow {\rm hadrons})
\approx |f|^2/\Gamma_R$ when $|\Delta m| < O(\Gamma_R)$. To obtain
$\Gamma(\psi(2S)\rightarrow R\rightarrow {\rm hadrons})\approx\Gamma(\psi(2S)
\rightarrow ggg)$, we need therefore  
\begin{equation}
    |f|^2 \approx \Gamma_R\Gamma(\psi(2S)\rightarrow ggg). 
                        \label{mixing}
\end{equation}
If $R$ is a light-quark resonance $q\overline{q}$, $|f|$ would be much too 
small for the following reason: While $|f|^2$ is of the order of 
$\Gamma(\psi(2S)\rightarrow ggg)\Gamma(R\rightarrow ggg)$ for 
$q\overline{q}$, we expect $\Gamma(R\rightarrow ggg)\ll \Gamma_R$ because of 
the $\alpha_s^3$ suppression of $q\overline{q}\rightarrow ggg$. 
Therefore there is no chance to satisfy Eq.(\ref{mixing}). It is likely
that the same argument applies to four-quark resonances. 

For glueballs, 
we simply do not have enough quantitative understanding to rule out a large 
enough coupling to $\psi(2S)$. Hou and Soni\cite{Hou} proposed a glueball near
the $J/\psi$ mass in order to enhance $J/\psi\rightarrow\rho\pi$ (and its
symmetry-related modes) but not other decay modes. To accomplish it, this
glueball must have very special, if not unnatural, properties\cite{BLT2}: 
It is nearly degenerate with $J/\psi$ with a quite narrow width for an object
of mass $\sim 3$ GeV and decays predominantly into $1^-0^-$.
Later Hou\cite{H} relaxed the constraint on the $\rho\pi$
branching to argue that such a glueball was not yet ruled out by
the search of the BES Collaboration\cite{BES2}.

In our case, since a glueball is introduced to account for the hadronic 
excess, it should couple not primarily to the $1^-0^-$ channels, but to many
other channels.  What we need is a generic vector glueball with no special
or unusual properties. If $\Gamma_R$ is as narrow as 100 MeV, for instance,
the mixing $|\varepsilon| = O(10^{-2})$ would be able to account for the 
excess in the inclusive hadron decay of $\psi(2S)$. The width can be wider.
In that case the pole transition strength $f$ should be stronger 
according to Eq.(\ref{mixing}).
Since the glueball $R$ couples to a photon only indirectly through its 
mixing to a quark pair, it is hard to detect $R$ in the hadronic 
cross section of $e^+e^-$ annihilation near the $\psi(2S)$ mass. 
Searching by hadronic reactions such as $p\overline{p}$ 
annihilation is a daunting task. From a purely experimental viewpoint, 
such a resonance has not been ruled out\cite{H}. 

Though the resonance scenario is admittedly a long shot, 
it is one of a very few options left to us.
One reason to pursue this somewhat unnatural scenario is that
the amplitude for $\psi(2S)\rightarrow R\rightarrow X$ has
automatically a large phase when $\Delta m < O(\Gamma_R)$ since 
the coupling $f$ is likely real, that is, dominated by the dispersive 
part. If that is the case, the resonant amplitude can interfere 
strongly with the three-gluon amplitude in two-meson decays.  
 
\section{Conclusion}
We have searched for a clue to solve the $\rho\pi$ puzzle in this paper. 
Our purpose is to locate the source of the problem rather than to offer
a final solution of the problem. Two threads have been exposed which 
may eventually lead us to a solution of the
$\rho\pi$ puzzle. They are the phases of the decay amplitudes and 
a possible excess in the inclusive hadronic decay rate of $\psi(2S)$. 
An experimental confirmation of the excess will be the most useful 
in directing theorists. If it is confirmed, it will be quite an
important experimental discovery by itself. One crucial experimental 
information will be the angular distributions of $J/\psi$ and 
$\psi(2S)$ into the channels other than $1^-0^-$ and $0^-0^-$. 
Difference in the angular distributions should have direct
correlation with enhancement and suppression in general.
As for the source of an additional process, the virtual 
$D\overline{D}$ pair and the vector glueball are two options that 
cannot be ruled out. To be frank, however, we admit that both ideas 
have unnaturalness.  More an attractive alternative is highly desired.
It is possible that the large relative phases so far observed in
$J/\psi$ decay are an accident and that the $\rho\pi$ puzzle is a problem 
of incalculable long-distance complications. However, 
our hope is that there might be something novel, simple, or fundamental 
hidden beneath the issue. 
  
\acknowledgements
The author thanks F.A. Harris and D. Paluselli for communications 
about the BES analysis and J.L. Rosner for a stimulating suggestion.
This work was supported in part by the Director, Office of Science, Office of
High Energy and Nuclear Physics, of the
U.S. Department of Energy under Contract DE--AC03--76SF00098 and in
part by the U.S. National Science Foundation under grant PHY--95--14797.

%%%%%%%%%%%%%%%%%%%%%%%%%%%%%%%
\begin{table}
\caption{Parametrization of the $\psi(2S)\rightarrow 1^-0^-$ amplitudes 
and values of branching fractions. The amplitude $\epsilon$ represents 
the $T_{33}$ breaking of $m_s-m_{ud}$ instead of $\lambda_8$ breaking. 
The $\eta$-$\eta'$ mixing angle is chosen to be $\theta_P = -20^{\circ}$. 
The fits to the central values are shown for the minimum and the maximum 
relative phase, $\delta = -\arg(a_1^*a_{\gamma})$= $0^{\circ}$ and
$\pm 90^{\circ}$. The ranges of values for $\rho\pi$ and $\omega\eta$ are
given for $|\epsilon/a_1| < 1/3$.}
\begin{tabular}{c|c|c|c|c}        
Modes & Amplitudes & Branchings (in $10^{-4}$) & 
\multicolumn{2}{c}{Fits}  \\ \cline{4-5}
  &  &  & $\delta= 0$ & $\delta= \pm 90^{\circ}$\\ \hline
$\rho^+\pi^-(=\rho^0\pi^0)$    & $a_1\;\;\;+\;\;\;  a_{\gamma}$ & $ <0.09 $ & 
        0$\sim$0.04 & 0.21$\sim$0.70 \\
$K^{*+}K^-$      & $a_1 + \epsilon + a_{\gamma}$& $<0.15 $ &
        0.00  & 0.30 \\
$K^{*0}\overline{K}^0$   
  & $a_1 + \epsilon -2a_{\gamma}$ & $0.41\pm 0.12\pm 0.08 $ &
        0.41 & 0.41  \\
$\omega\pi^0$ & $3a_{\gamma}$ & $0.38\pm 0.17\pm 0.11$ &
        0.38  & 0.38  \\
$\omega\eta$  & $\sqrt{1/3}(a_1+a_{\gamma})\;(\omega\eta_8)$ & $<0.33$ &
       $0.06\sim0.22$ & $0.05\sim 0.31$  \\
$\omega\eta'$ & $\sqrt{2/3}(b_1+b_{\gamma})\;(\omega\eta_1)$ & 
       $0.76\pm 0.44\pm 0.18$ & 0.76 & 0.76 \\
\end{tabular}
\label{table:1}
\end{table}

\begin{table}
\caption{Branching fractions of $J/\psi$ and $\psi(2S)$, and the
ratio $B(\psi(2S)\rightarrow h)/B(J/\psi\rightarrow h)$.
The $\pi^+\pi^-\pi^0$ mode is not included here since $J/\psi\rightarrow h$ 
is entirely $J/\psi\rightarrow\rho\pi$ within experimental uncertainties.}
\begin{tabular}{c|c|c|c}
Modes & Branchings of $J/\psi$ & Branchings of $\psi(2S)$ & 
     Ratio \\ \hline
$2(\pi^+\pi^-\pi^0)$ & $3.37 \pm 0.26 \times 10^{-2}$ & 
                 $3.0\pm 0.8\times 10^{-3}$ & $8.9 \pm 2.5$ \% \\
$3(\pi^+\pi^-)\pi^0$ & $2.9 \pm 0.6 \times 10^{-2}$  & 
                 $3.5 \pm 1.6\times 10^{-3}$ & $12.1 \pm 6.1$ \% \\
$K^+K^-$ & $2.37 \pm 0.31 \times 10^{-4}$ & $1.0 \pm 0.7 \times 10^{-4}$ &
                 $42 \pm 30$ \% \\
$\pi^+\pi^-K^+K^-$ & $7.2 \pm 2.3 \times 10^{-3}$ & $1.6 \pm 0.4 \times
              10^{-3}$ & $22.2 \pm 8.8$ \% \\
$p\overline{p}$ & $ 2.12 \pm 0.10 \times 10^{-3}$ & $1.9 \pm 0.5 \times 
              10^{-4}$ & $8.9 \pm 2.4$ \% \\
$p\overline{p}\pi^0$ & $1.09 \pm 0.09 \times 10^{-3}$ & $1.4 \pm 0.5 \times 
              10^{-4}$ & $12.8 \pm 4.7$ \% \\
$p\overline{p}\pi^+\pi^-$ & $6.0 \pm 0.5 \times 10^{-3}$ & $8 \pm 2 \times
              10^{-4}$ & $13.3 \pm 2.1$ \% \\
\end{tabular}
\label{table:2}
\end{table}

%%%%%%%%%%%%%%%%%%%%%%%%%%%%%%%%%%%%%%%%%%%%%%%%%%%%% 

%\input psfig
\noindent
\begin{figure}
\epsfig{file=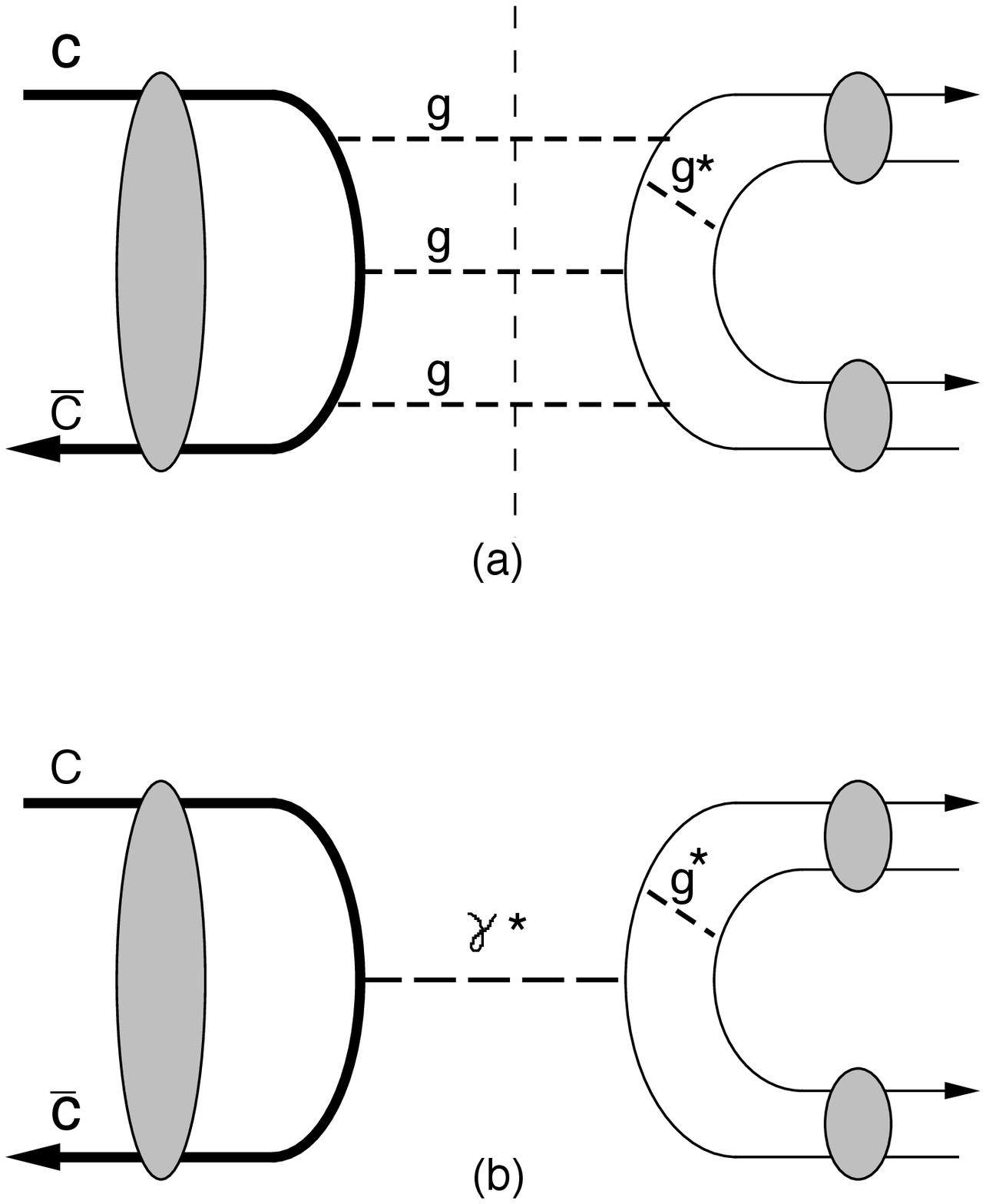,width=7cm,height=7.5cm}
\caption{Decays of charmonium into two mesons (a) through $ggg$
and (b) through one photon. The vertical broken line in diagram (a)
indicates that the gluons are placed on the mass shell when the inclusive
decay rate is computed with $ggg$.If perturbative QCD dominated, 
the one-gluon-exchange diagram depicted in (b) would dominate
in the final state of the one-photon process.
\label{fig:1}} 
\end{figure}

\noindent
\begin{figure}
\epsfig{file=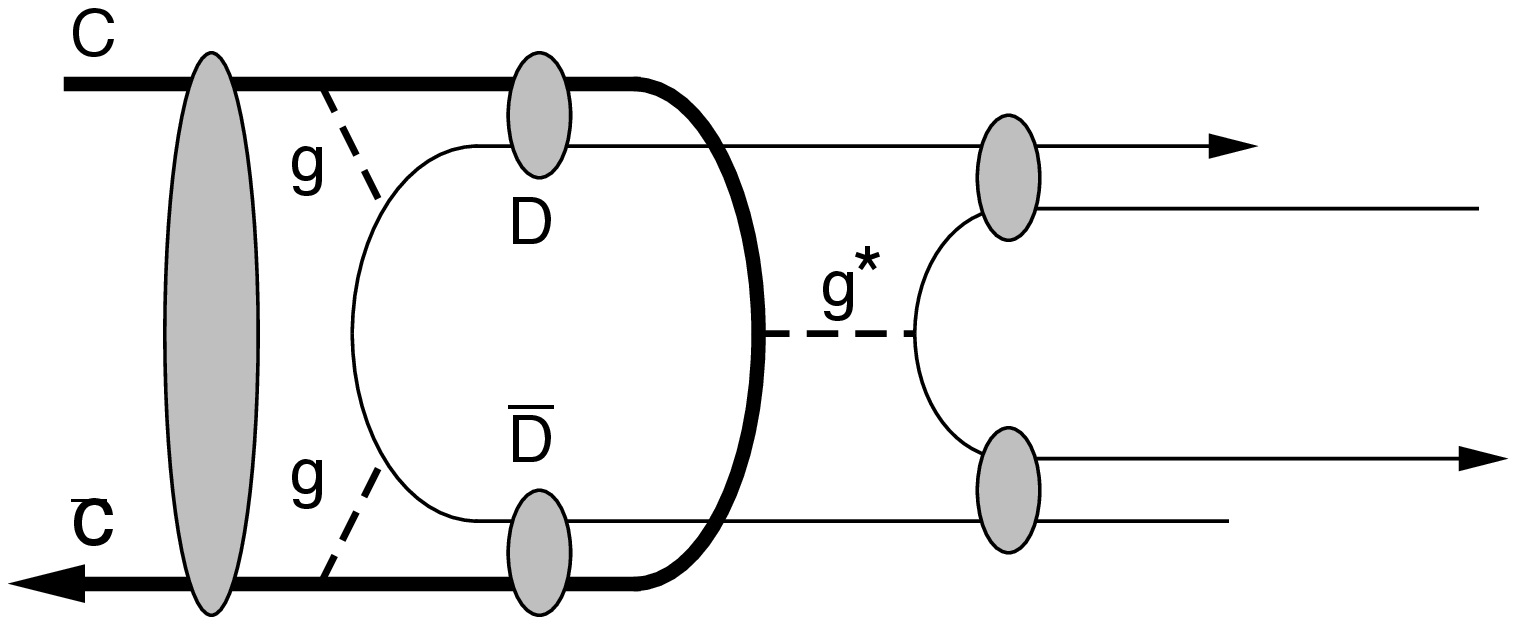,width=6cm,height=2.5cm}
\caption{The decay $\psi(2S)\rightarrow D\overline{D}$ (off shell)
$\rightarrow {\rm meson + meson}$.
\label{fig:2}} 
\end{figure}

\noindent
\begin{figure}
\epsfig{file=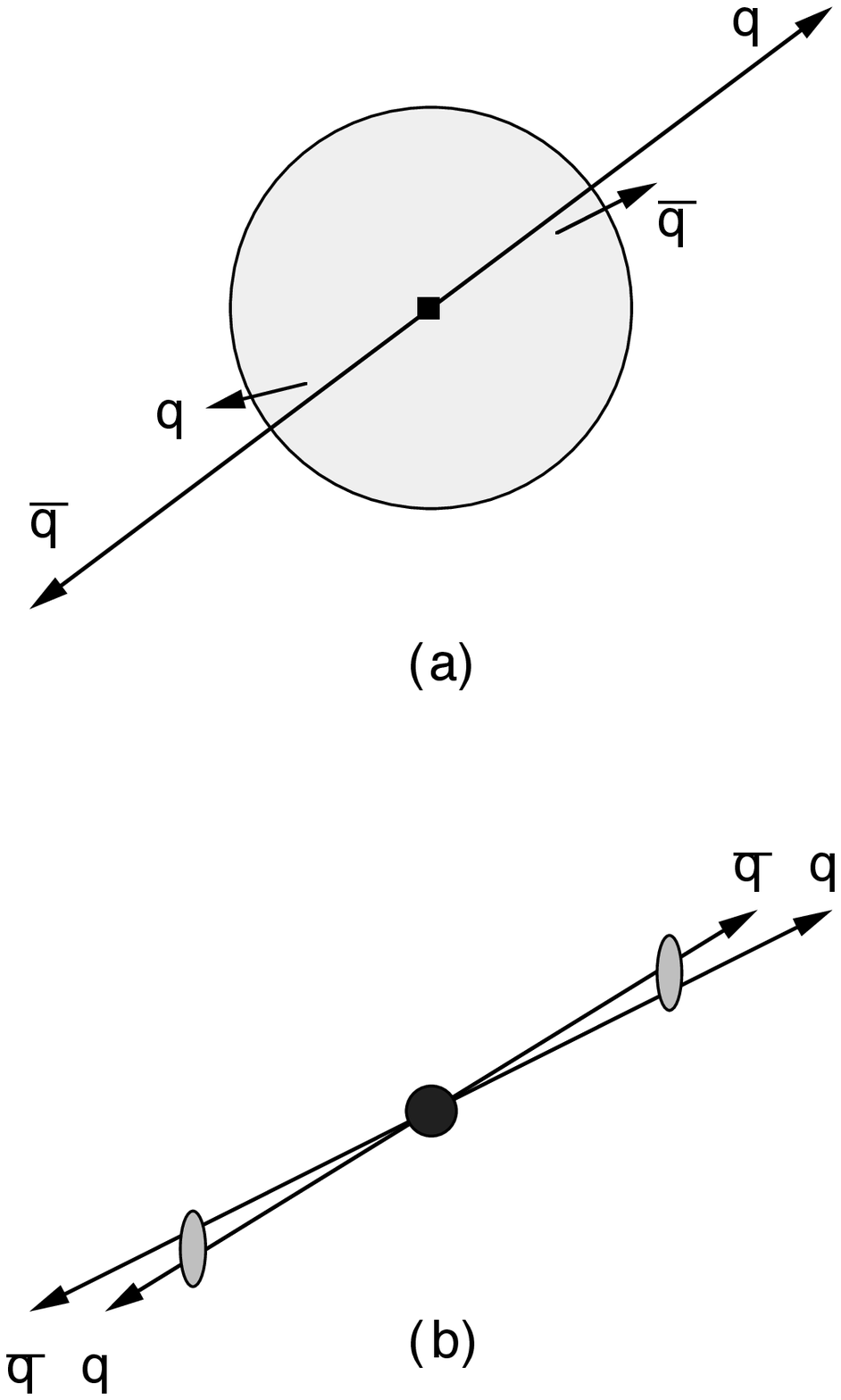,width=4cm, height=6.5cm}
\caption{(a) Formation of a light meson pair by energetic 
$q\overline{q}$ from $c\overline{c}$ and wee $q\overline{q}$ 
in $D\overline{D}$ annihilation. The invariant mass is small 
for $qq$ and for $\overline{q}\overline{q}$. The arrows denote 
directions and magnitudes of momenta. (b) Light meson pair
formation in one-photon annihilation where all quarks are hard.
\label{fig:3}}
\end{figure}
\end{document}